\RequirePackage{fix-cm}
\documentclass[twocolumn]{svjour3}          
\smartqed  
\usepackage{graphicx}
\usepackage{multirow}
\usepackage{booktabs}
\usepackage[misc]{ifsym}

\journalname{RDTM}
\begin{document}

\title{The design and implementation of GECAM satellite payload performance monitoring software
}



\author{
Peng Zhang\textsuperscript{1,3}
 \and Xiang Ma\textsuperscript{1,2*}
 \and Yue Huang\textsuperscript{1,2} 
 \and Shaolin Xiong\textsuperscript{1,2}
 \and Shijie Zheng\textsuperscript{1,2}
 \and Liming Song\textsuperscript{1,2}
 \and Ge Ou\textsuperscript{1,2}
 \and Yanqi Du\textsuperscript{3} 
 \and Jing Liang\textsuperscript{3} 
 \and Hong Wu\textsuperscript{3} 
}

\institute{
\Letter Xiang Ma\\
\email{max@ihep.ac.cn}\\         
\at
 {1} Institute of High Energy Physics, Chinese Academy of
Sciences, Beijing, 100049, China. 
 \at 
\\
 {2} University of Chinese Academy of Science, Beijing, 100049, China.
\at
\\
 {3} Southwest Jiaotong University, Chengdu, 610092, China.
\\
}


\date{Received: date / Accepted: date}

\maketitle

\begin{abstract}

\ \\
\textbf{Background}
The Gravitational wave high-energy Electromagnetic Counterpart All-sky 
Monitor (GECAM) is primarily designed to spot gamma-ray 
bursts corresponding to gravitational waves. 
In order to achieve stable observations from various astronomical phenomena, 
the payload performance need to be monitored during the in-orbit operation.
\\
\textbf{Method}
This article describes the design and implementation of 
GECAM satellite payload performance monitoring (GPPM) software. 
The software extracts the payload status and telescope observations 
(light curves, energy spectrums, characteristic peak 
fitting of energy spectrums, etc) from the payload data. 
Considering the large amount of payload status parameters 
in the engineering data, 
we have designed a method of parameter processing 
based on the configuration tables. 
This method can deal with the frequent changes of the data formats 
and facilitate program maintenance.
Payload status and performance are monitored through 
defined thresholds and monitoring reports. 
The entire software is implemented in python language
and the huge amount of observation data is stored in MongoDB.
\\
\textbf{Conclusion} 
The design and implementation of GPPM software have been completed, 
tested with ground and in-orbit payload data. The software 
can monitor the performance of GECAM payload effectively. 
The overall design of the software and the data processing 
method can be applied to other satellites.

\keywords{GECAM \and Payload \and Data process \and Python}
\end{abstract}

\section{Introduction}
\label{intro}
The first detection of gravitational wave signal was made on December 14, 2015 and 
was announced on February 11, 2016 \cite{GW 0}.
The direct detection of gravitational wave has started a revolutionary era of gravitational wave astronomy \cite{GW 1,GW 2}. 
Gravitational wave events occur randomly, and each event contains 
valuable information and significant opportunities for discovery.
The detection of high-energy electromagnetic counterparts of gravitational wave events 
as an important method for studying gravitational wave sources and 
fundamental physical laws has become the research hotspot of astronomy. 
\par
The Gravitational wave high-energy Electromagnetic Counterpart All-sky Monitor (GECAM) satellite 
is a full-sky monitoring telescope which can detect high-energy electromagnetic 
counterparts of gravitational wave events \cite{GECAM system design}. 
GECAM also monitors other burst events, such as 
gamma-ray bursts (GRBs), the high-energy radiation of fast radio bursts, and magnetar bursts.
GECAM measures the energy spectrum, light curves, and location of these burst events. 
GECAM is composed of two small satellites which operate in the low earth orbit. 
The two satellites operate in opposite orbital phase and opposite geocentric 
directions to obtain all-sky view, while each satellite has 25 Gamma-Ray Detectors (GRDs) 
and 8 Charged Particle Detectors (CPDs) \cite{GECAM payload}. 
GRDs are applied to detect gamma-rays in the energy range from 6 keV to 5 MeV. 
GRDs are comprised of {$\rm LaBr_{3}$}:Ce scintillator, SiPM array 
and preamplifier; the large dynamic energy range of GRDs are achieved by 
the high-gain and low-gain ADC channels of the preamplifier \cite{GECAM GRD 1,GECAM GRD 2}. 
The location of GRB is calculated from the photon counts from 
different pointing modes of GRDs \cite{GECAM localization method}. 
CPDs are insensitive to gamma rays but are designed to detect energetic charged 
particles of energies from 300 keV to 5 MeV. CPDs can be used to exclude interference 
from space charged particle bursts during data analysis. 
The telescope observation data and payload status (voltage, current, temperature, etc) of the GRDs 
and CPDs are processed by five data acquisition devices (DAQs).

\par
The GECAM has been successfully launched in December 2020. The GECAM mission consists of six systems: 
Satellite System; Launch Vehicle System; Launch Site System; Tracking, Telemetry and Command System (TT\&C);
Ground Support System and Scientific Application System. 
The Ground Support System is developed by National Space Science Center, Chinese Academy of Sciences (NSSC), 
which receives, pre-processes, manages and archives satellite data. 
Scientific Application System links scientific research and satellite operations, 
which monitors the performance of the payload to ensure the stable operation 
of the instruments and implements the calibration, production, 
storage, distribution and rapid processing of scientific data \cite{GECAM sci system,GECAM data analysis}. 
GECAM satellite payload performance monitoring (GPPM) software 
is a part of Scientific Application System. 
The software firstly process payload raw data to generate data products, 
and then analyse the parameters of the data products to monitor the payload performance.
When the satellite is in orbit, the raw data processing is 
implemented by the Ground Support System. 
The raw data processing method processes the raw data which generated from payload ground test. 
Payload raw data is unpacked and reorganized to the data products which 
are used for functional and interface testing by the various 
subsystems of Scientific Application System. 
\par
In this paper, we describes the process of designing and 
implementing of the GPPM software. 
First, general information and structure of the software are presented.
Second, the detailed implementation of each module of the software is described.
Finally, the testing process for the functionality of the software is illustrated.

\section{Software overview}
\label{sec:software overview}
\begin{figure}[htbp]
\centering
  \includegraphics[width=0.5\textwidth]{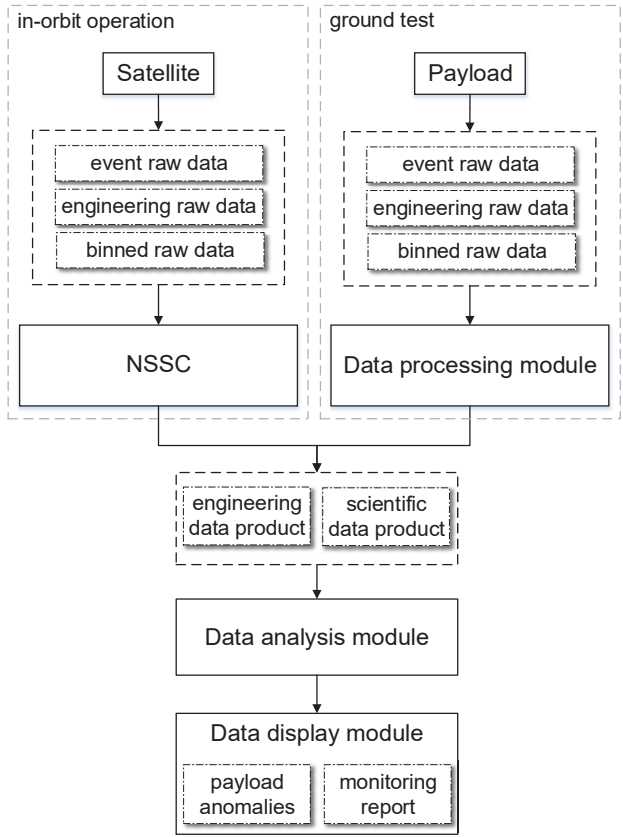}
\caption{The flowchart of the GPPM software}
\label{fig:flowchart system}       
\end{figure}

The main task of GPPM software is to monitor the performance of payload. 
The flowchart of the software is shown in Figure. \ref{fig:flowchart system}. 
The software consists of three modules: data processing 
module, data analysis module and data display module.
When satellite is in ground test, the data processing module of this software 
unpacks and reorganizes the payload raw data to level 0 data product. 
During in orbit performing, the data processing is implemented by the NSSC. 
The correspondence between raw data and data product is shown in Table. \ref{tab:data relationship}.
Data analysis module extracts payload status and telescope 
observations from payload data, then determines anomalies from the 
thresholds for crucial monitoring objects, and finally stores 
the observations and anomalies in the database. 
User analyzes the payload performance from anomalies and 
monitoring reports in the data display module.
The software uses 17 and 582 data tables in MySQL \cite{mysql introduction} and 
MongoDB \cite{mongodb introduction} for data storage, respectively. 
Regular data is stored in MySQL, such as lists of files, thresholds for anomalies, 
anomalies informations, etc. 
The observations data are stored in MongoDB. 
MongoDB is more efficient than MySQL when querying for 
large amounts of data \cite{mysql mongodb compare}. 
From satellite launch to July 2021, MongoDB have stored 124 GByte of data.
The whole software is implemented in python 3.6 \cite{python introduction}.
The functions of the three modules of the software are described below:
\par
Data processing module unpacks the payload raw data according 
to pre-defined format. 
The parameters in the raw data are converted to physical values according to 
calculation formula and saved to the data product.
While the satellite's payload had been tested on the ground, 
payload data management software generated large number of 
scientific and engineering raw data for payload tests. 
The data processing module unpacks, converts and reorganizes these payload raw data 
to produce data products for analysing the performance of the payload. 
When satellite operates in orbit, satellite transmits 
scientific and engineering raw data via telemetry and data transmission channels. 
When Ground Support System receives the payload raw data, 
it processes these raw data to data products and transfers to Scientific Application System 
via virtual private network (VPN) for further processing. 
\par
The function of data analysis module is 
to extract the payload status and telescope observation 
data (such as light curve and energy spectrum) 
of each detector from the data products. 
For crucial monitoring objects, this module needs 
to decide whether the objects is abnormal based on the defined threshold.
\par
The data display module generates payload performance monitoring reports 
for the past period at regular intervals and creates web pages 
for viewing historical data, anomalies, monitoring reports etc. 

\begin{table*}[htbp]
\caption{Correspondence between processed raw data and generated data product}
\label{tab:data relationship}  
\begin{tabular}{llll}
\toprule
Processed raw data & Generate & Corresponding data product   & Category of data product    \\
\noalign{\smallskip}\hline\noalign{\smallskip}
\multicolumn{1}{c}{\multirow{2}{*}{Engineering raw data}} & \multirow{2}{*}{ $\rightarrow$} & Payload engineering data product & \multirow{2}{*}{Engineering data product} \\
\multicolumn{1}{c}{}  &   & Orbital data product    &         \\
\noalign{\smallskip}\hline\noalign{\smallskip}
Event raw data       & \multirow{2}{*}{ $\rightarrow$  } & Event data product      & \multirow{2}{*}{Scientific data product}  \\
Binned raw data       &      & Binned data product          &          \\                                
\bottomrule
\end{tabular}
\end{table*}

\section{Module design}
\label{sec:Module design}

\subsection{Data processing module}
\label{sec:data process}
The data processing module queries the binary raw data required to process 
from the MySQL for every ten minutes, and multi-threads unpacks the raw 
data files in parallel to generate data products 
in FITS\footnote{FITS format is a common data storage 
format in the astronomical field, for details see the web 
site of https://fits.gsfc.nasa.gov/} format. 
The binary raw data consists of several packages. 
The length of each package is 884 bytes; the first 12 bytes 
contain information including the start flag and the 
type of package; the rest 872 bytes are the data field. 
Raw data are divided into three types: event raw data, 
engineering raw data and binned raw data.
\par
The data field of each package in the event raw data consists of 
the DAQ number (2 bytes) and 145 events (the length of each event is 6 bytes). 
The arrival time of each event is recorded using a local time 
counter with a time resolution of 0.1 microsecond. 
There are six types of events in total.
Physical events contain time, energy channel (4096 channels in total) and dead time 
(time consumed by DAQ to process a single photon) of the photon.
Other types of events are recorded at one second intervals to aid 
in the calculation of full arrival time of each photon. 
Arrival time, energy, dead time and information used to assist 
with time calculation of each photon are stored in the event data product.
\par
The binned raw data consists of time binned raw data 
and energy binned raw data. The time binned raw data records 
the photon counts and sum of dead time every 50 milliseconds.
The energy binned raw data records photon counts and sum of dead time per 
second, the photon counts are the number of photons 
in each channel after the original 4096 energy channels have been merged into 128 channels. 
The photon counts and dead time information extracted 
from time binned raw data and energy binned raw data are saved as 
time binned data product and energy binned data product, respectively.
\par
Engineering raw data records information of payload 
status every second, such as temperature, current, voltage 
and photon counts (both processed and downlinked) of the detectors. 
The satellite's attitude orbit information is also recorded in engineering raw data.
The data field of each package from the engineering raw data 
consists of 511 parameters, while the individual parameters are stored by one or more bits. 
We need to extract the original value of the parameter based on 
the number of bits in the corresponding position of the package, 
and then convert it to the physical value using the defined formula. 
The status of the payload is saved to payload engineering data product. 
The attitude and orbit data of the satellite are saved to orbital data product.
For the engineering raw data that contains large amount of parameters 
needs to be processed, we have designed a method of parameter processing 
according to the configuration table. 
If changes are made to data format, parameter processing can be quickly 
updated by adjusting the configuration table 
directly without modifying the program code. 
The flowchart of the method is shown in 
the Figure \ref{fig:engineering raw data process}. 
The specific process flow is described below:

\begin{figure}[htbp]
\centering
  \includegraphics[width=0.2\textwidth]{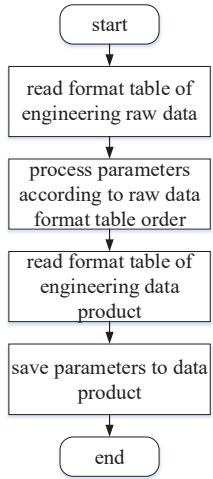}
\caption{The flowchart of the engineering raw data processing}
\label{fig:engineering raw data process}       
\end{figure}

\begin{enumerate}
\item The first step is to read the description table of engineering raw data 
format file which is stored by the CSV\footnote{CSV 
files store tabular data (numbers and text) in plain text form} file, 
see Table. \ref{tab:origin data format} for an example. 
The first and second columns represent the parameter name 
and the number of bits occupied by the parameters, respectively. The 
third column represents the type of data that the parameter 
needs to convert. The last column represents the formula 
required for the calculation after the parameter is converted.

\item The second step is to read the binary corresponding bits according to the 
second column, and convert it to the data type described in 
the third column. The final value is obtained according to 
the calculation formula of the last column. If the formula 
is not empty, it is converted to expression execution using 
python's built-in eval function. When the calculation formula 
of some parameters is complex and cannot be described by 
string, we need to pre-define its calculation function in 
the program and start with the character formula as the function name.

\item The third step is to read the description table of engineering data product 
format table which is also stored in the CSV file, see 
Table. \ref{tab:data product format} for an example. Data products 
in FITS format use several Header Data Units (HDUs) to 
distinguish data categories. We choose 
to store data in HDUs using data arrays. The second and third 
columns in Table. \ref{tab:data product format} indicate the 
name and format of each column in the data array, respectively. 
The fourth column indicates the unit of data. The last column indicates that the program 
can get the corresponding parameter value from the engineering raw data 
via the parameter name.

\item Finally, the values corresponding to parameters are 
obtained from engineering raw data according to Table. \ref{tab:data product format}, and 
the descriptions in this table are further used to generate engineering 
data products in the corresponding format.
\end{enumerate}

\begin{table*}[htbp]
\caption{Example of engineering raw data parsing format table}
\label{tab:origin data format}       
\begin{tabular}{llll}
\toprule
Parameter name & Bit & Parameter type & Calculation  \\
\noalign{\smallskip}\hline\noalign{\smallskip}
Telemetry package identification & 16 & Hexadecimal & - \\
Local accumulation of seconds of pulse time & 40 & Binary & formula\_local\_time(value) \\
Receive command count & 8 & Decimal & - \\
Physical Package Delivery Count & 8 & Decimal & value*1024 \\
+30V voltage & 8 & Decimal & (value/255)*2*1050/50 \\
Temperature sensitive resistor telemetry of GRDA & 12 & Decimal & formula\_warm\_resistance(value) \\
Digital temperature telemetry of GRDA & 12 & Decimal & formula\_digital\_temperature(value) \\
Rate meter for high gain processing data of GRDA & 16 & Decimal & value*4 \\
Rate meter for high gain downlink data of GRDA & 16 & Decimal & value*4 \\
Rate meter for high low processing data of GRDA & 16 & Decimal & value*4 \\
Rate meter for high low downlink data of GRDA & 16 & Decimal & value*4 \\

J2000 coordinate system position X & 32 & Float & - \\
J2000 coordinate system position Y & 32 & Float & - \\
J2000 coordinate system position Z & 32 & Float & - \\
\bottomrule
\end{tabular}
\end{table*}

\begin{table*}[htbp]
\caption{Example of the specification table of engineering data product format}
\label{tab:data product format}      
\begin{tabular}{lllll}
\toprule
&&&&\multicolumn{1}{l}{Parameter of raw data from Table. \ref{tab:origin data format} }  \\ \cmidrule{5-5}
 HDU name & Data name & Data format & Data unit & Parameter name  \\
\noalign{\smallskip}\hline\noalign{\smallskip}
\multirow{7}{*}{GRD01}& TIME & D & s & Local accumulation of seconds of pulse time \\
& GRDA\_Resist & E & - & Temperature sensitive resistor telemetry of GRDA \\
& GRDA\_Temp & E & degree & Digital temperature telemetry of GRDA \\
& GRDA\_HGainCnt1 & J & - & Rate meter for high gain processing data of GRDA \\
& GRDA\_HGainCnt2 & J & - & Rate meter for high gain downlink data of GRDA \\
& GRDA\_LGainCnt1 & J & - & Rate meter for low gain processing data of GRDA \\
& GRDA\_LGainCnt2 & J & - & Rate meter for low gain downlink data of GRDA \\
\bottomrule
\end{tabular}
\end{table*}

\begin{figure}[htbp]
\centering
  \includegraphics[width=0.35\textwidth]{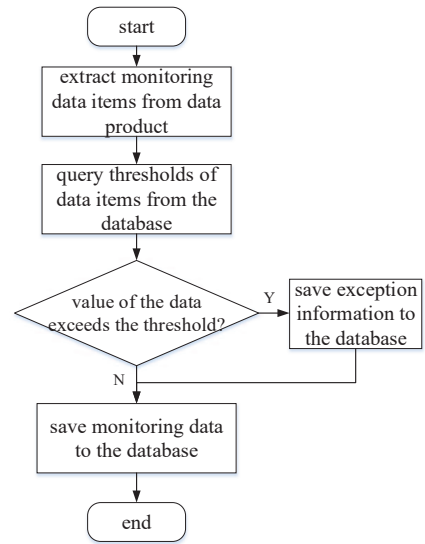}
\caption{The flowchart of the data analysis module}
\label{fig:data product processing}       
\end{figure}

\begin{table*}[htbp]
\caption{Example of engineering parameter table in mongoDB database}
\label{tab:engineering mongo format}     
\begin{tabular}{lllll}
\toprule
&&&\multicolumn{2}{l}{Data of data product from Table. \ref{tab:data product format} } \\ \cmidrule{4-5}

Table name & Column name  & Column type  & HDU name & Data name \\
\noalign{\smallskip}\hline\noalign{\smallskip}
\multirow{5}{*}{dataMonitor\_a\_grdgcllj} & \_id & Integer & GRD01 & TIME\\
& grd1\_high\_rate & Integer & GRD01 & GRDA\_HGainCnt1\\
& grd2\_high\_rate & Integer & GRD02 & GRDB\_HGainCnt1\\
& grd3\_high\_rate & Integer & GRD03 & GRDC\_HGainCnt1\\
& grd4\_high\_rate & Integer & GRD04 & GRDD\_HGainCnt1\\
& grd5\_high\_rate & Integer & GRD05 & GRDE\_HGainCnt1\\
\bottomrule
\end{tabular}
\end{table*}

\subsection{Data analysis module}
\label{sec:data analysis module}
Data analysis module regularly queries whether there are 
unprocessed data products in MySQL every ten minutes. 
Current module is multi-threaded to process data products, 
and save the required observation data to the MongoDB. 
Meanwhile, this module determines whether the data is in an 
abnormal state using defined thresholds and meanwhile saves the abnormal 
information to MySQL. 
The range of exception thresholds corresponding to the data 
values is also stored in the MySQL.
The flowchart of data analysis module is shown in Figure \ref{fig:data product processing}. 
\par
Scientific data product includes event data product and binned data product.
For the scientific data product, we extract the light 
curve (time scale is one second) and energy spectrum of 
each detector from event data product and binned data product.
The sum of the dead times per second is extracted from time binned data product.
We determine detector anomalies by whether the photon counts 
in the light curve exceed the defined threshold.
In order to measure the performance of the detector in several dimensions, 
we extract the light curves of electronic noise (channel < 200) 
and extract light curves of the split channels 
(channel < 300, 300 $ \sim $ 1000, 1000 $ \sim $ 3000 and > 3000) 
from event data product; the time scale of these light curves are 1 second.
Gaussian function is applied fit the characteristic peak (37.4 KeV, 511 KeV and 1470 KeV) 
in the energy spectrum, while the shift in the characteristic peak reflects the change in 
the detector's energy response\footnote{The 37.4 keV and 1470 keV intrinsic 
gamma-ray lines of {$\rm LaBr_{3}$}:Ce can be resolved from in-flight backgrounds. 
The galactic 511keV gamma-ray line also can be resolved \cite{GECAM GRD 2}}.
Counts of noisy light curve and fitting parameters of the characteristic peaks 
that exceeded the threshold are considered anomalies. 
The extracted data and anomalies will be saved in the database. 
\par 
The engineering data product includes orbital data product and 
payload engineering data product.
The satellite's orbit, attitude, time and speed per second are stored in orbital data product. 
The angle between payload z-axis and geocentric are calculated 
based on the satellite's attitude and orbit. 
From the satellite's attitude, orbit and time information, 
we calculate the angle between the normal direction of satellite's solar panel 
and solar incidence, and whether the satellite is in the solar irradiation zone.
The payload engineering data product stores the status 
parameters of each detector such as electric current, voltage, temperature, etc. 
We need to extract the corresponding 
values of each parameter every second and determine 
whether the parameter is in an abnormal state based on the defined threshold. 
Since the 536 parameters in the payload engineering data product 
need to be stored using 144 MongoDB data tables, 
we have also designed a method for automatic engineering 
parameter extraction and saving based on configuration tables. 
We first extract the corresponding values from the 
payload engineering data products according to the parameters 
in Table. \ref{tab:data product format} and determine the exceptions 
according to the defined thresholds. 
We then obtain the corresponding engineering parameters from 
the engineering data product according to the HDU name and data name 
in the format of the MongoDB data table, the format is shown on 
Table. \ref{tab:engineering mongo format}. 
Lastly, the parameters are saved in the corresponding data table.

\subsection{Data display module}
In order to facilitate users to view data, we build a web system 
to provide data query and display. 
We can view all the historical observation data of the telescope 
and the abnormal payload information found in the data analysis module.
This module generates monitoring reports at regular intervals 
for the past period (the exact time period can be modified in this module) 
based on the data stored in the MongoDB. 
The monitoring report contains status parameters for all detectors 
such as temperature, bias voltage, baseline, etc. The total photon 
counts, the photon counts in different energy ranges, energy spectrum, 
dead time, and the satellite position per second 
are also shown in the monitoring report. 
The monitoring report also contains overall information of 
the satellite such as attitude, orbit, velocity, position, etc.
Payload performance can be analysed by anomaly records and monitoring reports.
\par
The operation of different users 
can be controlled by permissions in this module. 
All users can view the observation data of telescope 
at different time intervals, such as the light curve, 
the energy spectrum and the fitted parameters of the characteristic 
peaks of the energy spectrum.
Administrators will authorise users, modify the 
thresholds for determining parameter anomalies in the 
data product analysis module, and the time 
for generating monitoring reports. 

\section{Software test}
\label{sec:software test}
\begin{figure*}[htbp]
\centering
  \includegraphics[width=0.8\textwidth]{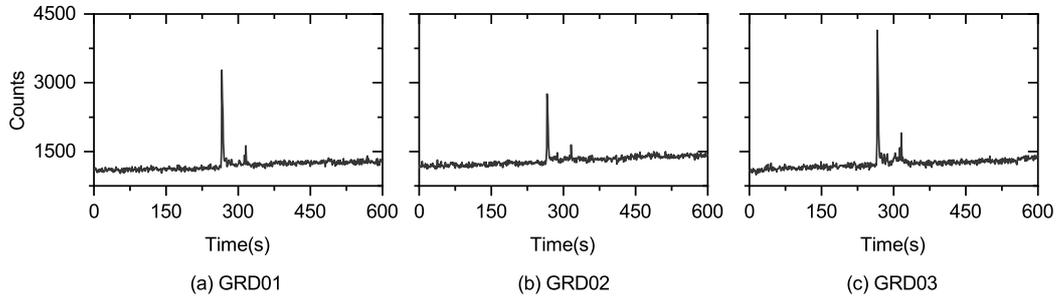}
\caption{Light curves of the three GRD with one second time bin. 
Each light curve is extracted from the whole energy band, and the time is relative to UTC 2021-06-19T23:55:00.}
\label{fig:light curve}      
\end{figure*}

\begin{figure*}[htbp]
\centering
 \includegraphics[width=0.8\textwidth]{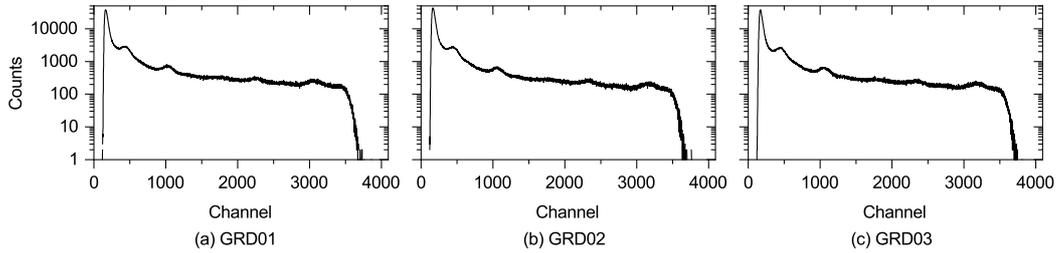}
\caption{Energy spectrums of the three GRD.  The time range for each 
energy spectrum is from UTC time 2021-06-19T22:00:00 to 2021-06-19T23:00:00.}
\label{fig:energy spectrum}       
\end{figure*}

\begin{figure*}[htbp]
\centering
 \includegraphics[width=1\textwidth]{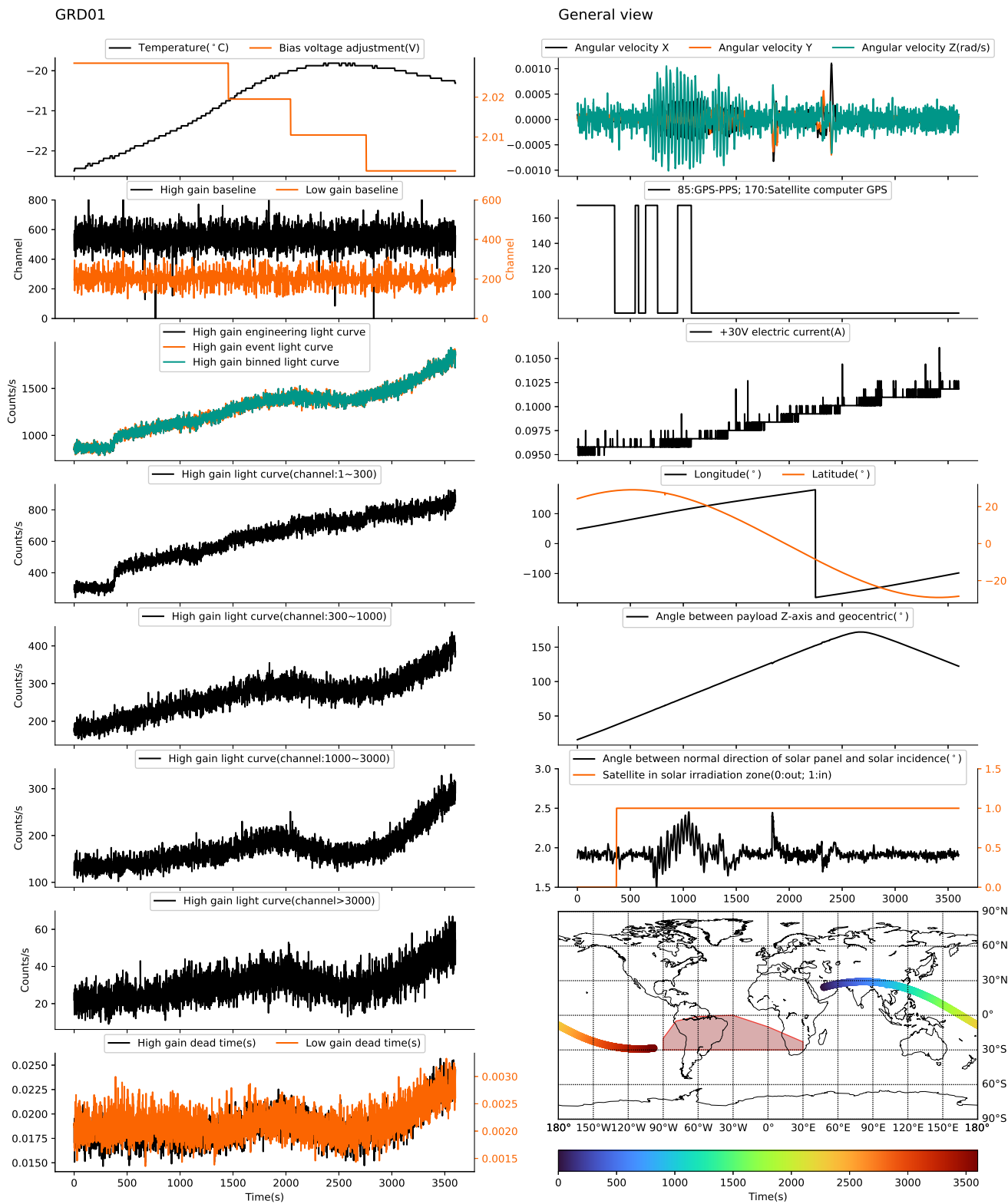}
\caption{Example of payload performance monitoring report. The figures in the left column  
represent some monitoring objects of the 
GRD01 detector, from top to bottom are temperature and 
bias voltage adjustment, baseline, light curves, light curve in 
different channel ranges, dead time, respectively.
The figures in the right column show the monitoring objects 
of the overall satellite status, from top to bottom are angular velocity, 
source of the time signal that marks the second, +30V electric current, 
longitude and latitude, angle between payload Z-axis and geocentric, 
satellite charging information, trajectory of satellite 
(gradient colour represents time change, and red area indicates the South Atlantic Anomaly), respectively. }
\label{fig:monitoring report}       
\end{figure*}

The data processing module unpacks and reorganizes
the binary raw data to data product during the payload 
ground test, and this module is involved in experiments 
such as mechanics experiments, temperature experiments, load experiments, etc. 
Data analysis module extracts payload status and telescope observations from data products, 
and determines anomalies based on defined thresholds. 
The observation data and anomalies are used to 
analyse how the payload performs. 
Figure \ref{fig:light curve} shows the light curves of a 
GRB event GRB210619B\footnote{https://gcn.gsfc.nasa.gov/gcn3/30264.gcn3} 
that was detected by GECAM.
Figure \ref{fig:energy spectrum} gives an example of energy spectrum. 
The monitoring reports generated by the data display 
module at regular intervals are shown in the Figure \ref{fig:monitoring report}. 
We can intuitively analyse the impact of the detector's 
status on the observation performance from the monitoring reports. 
The light curves, energy spectrums and payload monitoring report 
are consistent with the theoretical expectations. 
\par
Until July 2021, the satellite has been tested in orbit for 7 months. 
We have counted the number of various data products processed 
during this period, as shown in Table \ref{tab:statistics in-orbit}. 
We test the performance of the software in processing one hour of data products.
The CPU used in both testing and operating environment is 
an Intel(R) Xeon(R) Silver 4210 with frequency of 2.2 GHz. 
Performance test results are shown in Table \ref{tab:process data product consuming}, 
and the processing time for each category of data product is 
less than the 5 minutes required by the software design.
\par
Tests using ground tests and in-orbit operational data have 
demonstrated that the software functions correctly. 
Performance tests show that the software's performance meets 
the software's design requirements.

\section{Conclusion}
\label{sec:conclusion}
The GPPM software is designed to monitor the performance of GECAM payload. 
The software unpacks and reorganizes the payload raw data to data product. 
Payload status and telescope observations from data product 
are extracted and saved to MongoDB. 
Payload anomalies are obtained by testing whether 
the extracted data exceeds defined threshold value. 
Considering that the large number of parameters need to be processed in engineering 
raw data and data product, we have designed a method of processing 
parameters according to configuration table which makes the program easier to maintain.
Regularly generated monitoring reports based on historical data in 
the database can be used to analyse correlation of payload parameters and anomalies.
\par
Tested with ground and in-orbit payload data, the functionality and performance 
of the software meets the design requirements. 
The software enables the performance of the payload to 
be monitored effectively and guarantees the proper functioning of GECAM. 
The software is more flexible and practical, and can be applied to other 
satellite projects with appropriate modifications.

\begin{acknowledgements}
This work is supported by the Strategic Priority Research Program of Chinese 
Academy of Sciences (Grant Nos. XDA15360300, XDA15360302 and XDB23040400).
\end{acknowledgements}

%
%


\begin{table*}[htbp]
\caption{Statistics on the number of data products processed during the in-orbit testing}
\label{tab:statistics in-orbit}       
\begin{tabular}{llll}
\toprule
Category & Data product & Total file number & Total file size(Mbyte)  \\
\noalign{\smallskip}\hline\noalign{\smallskip}
\multirow{2}{*}{Engineering}  & Payload engineering data product & 5729 & 19451 \\
& Orbital data product & 4918 & 916 \\
\multirow{2}{*}{Scientific} & Event data product & 7852 & 4696045 \\
 & Binned data product & 11187 & 551818 \\
\bottomrule
\end{tabular}
\end{table*}

\begin{table*}[htbp]
\caption{Performance of data product processing}
\label{tab:process data product consuming}       
\begin{tabular}{llll}
\toprule
Category & Data product & File size & Time consuming(s)  \\
\noalign{\smallskip}\hline\noalign{\smallskip}
\multirow{2}{*}{Engineering}  & Payload engineering data product & 4 MByte & 16.98 \\
& Orbital data product & 217 KByte & 20.38 \\
\multirow{2}{*}{Scientific} & Event data product & 958 MByte & 70.57 \\
 & Binned data product & 60 MByte & 32.20 \\
\bottomrule
\end{tabular}
\end{table*}

\bibliographystyle{paper}
\bibliography{{paper}}

\begin{thebibliography}{1}

\bibitem{dbhat}
D.~Bhat and S.~Nayar.
\newblock Ordinal measures for image correspondence.
\newblock {\em IEEE Transactions on Pattern Analysis and Machine Intelligence},
  20(4):415--423, 1998.

\bibitem{mongodb}
Cornelia Gy{\H{o}}r{\"o}di, Robert Gy{\H{o}}r{\"o}di, George Pecherle, and
  Andrada Olah.
\newblock A comparative study: Mongodb vs. mysql.
\newblock In {\em 2015 13th International Conference on Engineering of Modern
  Electric Systems (EMES)}, pages 1--6. IEEE, 2015.

\bibitem{liao2020localization}
Jin-Yuan LIAO, Qi~LUO, Yue ZHU, Xin-Ying SONG, Wen-Xi PENG, Shuo XIAO, Gang LI,
  and Shao-Lin XIONG.
\newblock The localization method of gecam and simulation analysis.
\newblock {\em SCIENTIA SINICA Physica, Mechanica \& Astronomica},
  50(12):129510, 2020.

\bibitem{zhang2019energy}
Dali Zhang, Xinqiao Li, Shaolin Xiong, Yanguo Li, Xilei Sun, Zhenghua An,
  Yanbing Xu, Yue Zhu, Wenxi Peng, Huanyu Wang, et~al.
\newblock Energy response of gecam gamma-ray detector based on labr3: Ce and
  sipm array.
\newblock {\em Nuclear Instruments and Methods in Physics Research Section A:
  Accelerators, Spectrometers, Detectors and Associated Equipment}, 921:8--13,
  2019.

\end{thebibliography}


\begin{thebibliography}{}
\bibitem{GW 0}
Abbott B P, Abbott R, Abbott T D, et al. Observation of gravitational waves from a binary black hole merger[J]. Physical review letters, 2016, 116(6): 061102.

\bibitem{GW 1}
Mészáros P, Fox D B, Hanna C, et al. Multi-messenger astrophysics[J]. Nature Reviews Physics, 2019, 1(10): 585-599.

\bibitem{GW 2}
Mostafá M. The Astrophysical Multi-messenger Observatory Network[J]. Nature Reviews Physics, 2020, 2(9): 446-448.


\bibitem{GECAM system design}
HAN X B, ZHANG K K, HUANG J, et al. GECAM satellite system design and technological characteristic[J]. SCIENTIA SINICA Physica, Mechanica \& Astronomica, 2020, 50(12): 129507.

\bibitem{GECAM payload}
LI X Q, Wen X, AN Z H, et al. The GECAM and its payload[J]. SCIENTIA SINICA Physica, Mechanica \& Astronomica, 2020, 50(12): 129508.


\bibitem{GECAM GRD 1}
Lv P, Xiong S L, Sun X L, et al. A novel gamma-ray detector for GECAM[C]//International conference on Technology and Instrumentation in Particle Physics. Springer, Singapore, 2017: 3-7.

\bibitem{GECAM GRD 2}
Zhang D, Li X, Xiong S, et al. Energy response of GECAM gamma-ray detector based on LaBr3: Ce and SiPM array[J]. Nuclear Instruments and Methods in Physics Research Section A: Accelerators, Spectrometers, Detectors and Associated Equipment, 2019, 921: 8-13.

\bibitem{GECAM localization method}
LIAO J Y, LUO Q, ZHU Y, et al. The localization method of GECAM and simulation analysis[J]. SCIENTIA SINICA Physica, Mechanica \& Astronomica, 2020, 50(12): 129510.

\bibitem{GECAM sci system}
Chen W, SONG L M, ZHENG S J, et al. Introduction of the scientific application system of GECAM[J]. SCIENTIA SINICA Physica, Mechanica \& Astronomica, 2020, 50(12): 129512.

\bibitem{GECAM data analysis}
SONG X Y, XIONG S L, LUO Q, et al. Introduction to gamma-ray burst data analysis algorithm and software tools for GECAM[J]. SCIENTIA SINICA Physica, Mechanica \& Astronomica, 2020, 50(12): 129511.

\bibitem{mysql introduction}
Official website of Mysql. https://www.mysql.com/

\bibitem{mongodb introduction}
Official website of Mongodb. https://www.mongodb.com/

\bibitem{mysql mongodb compare}
Jose B, Abraham S. Performance analysis of NoSQL and relational databases with MongoDB and MySQL[J]. Materials today: PROCEEDINGS, 2020, 24: 2036-2043.

\bibitem{python introduction}
Millman K J, Aivazis M. Python for scientists and engineers[J]. Computing in Science \& Engineering, 2011, 13(2): 9-12.



\end{thebibliography}

\end{document}